\documentclass[onecolumn,preprintnumbers,smaller,
superscriptaddress]{revtex4} 

\usepackage{amssymb,amsmath,graphicx,dcolumn,bm,epsfig,float}
\usepackage[utf8]{inputenc}
\usepackage[T1]{fontenc}
\usepackage[colorlinks,linkcolor=black,citecolor=blue]{hyperref}

\begin{document}

\title{ Low-power phonon lasing through position-modulated Kerr-type nonlinearity}
\author{P. Djorwe}
\email{philippe.djorwe@univ-lille1.fr}
\affiliation{Institut d’Electronique, de Microélectronique et Nanotechnologie, UMR
CNRS 8520 Université de Lille, Sciences et technologies, Villeneuve d’ Ascq 59652, France}

\author{Y. Pennec}
\email{yan.pennec@univ-lille1.fr}
\affiliation{Institut d’Electronique, de Microélectronique et Nanotechnologie, UMR
CNRS 8520 Université de Lille, Sciences et technologies, Villeneuve d’ Ascq 59652, France}

\author{B. Djafari-Rouhani}
\email{bahram.djafari-rouhani@univ-lille1.fr}
\affiliation{Institut d’Electronique, de Microélectronique et Nanotechnologie, UMR
CNRS 8520 Université de Lille, Sciences et technologies, Villeneuve d’ Ascq 59652, France}

\begin{abstract}
We demonstrate low-power amplification process in cavity optomechanics (COM). This operation is based 
on the nonlinear position-modulated self-Kerr interaction. Owing to this nonlinear term, the effective 
coupling highly scales with the photon number, resulting in a giant enhancement of the cooperativity. 
Even for small nonlinearity, the system reaches the amplification threshold for weak driving strength, 
leading to low-power phonon lasing. This amplifier can be phase-preserving and provides  a
practical advantage related to the power consumption issues. This work opens up new avenues to 
realize low-power and efficient amplifiers in optomechanics and related fields.  

\end{abstract}

\maketitle

\date{\today}

\section{Introduction}

Cavity optomechanics (COM), which is devoted to explore interaction between electromagnetic radiation 
and mechanical object, provides a platform to perform phonon lasing action \cite{[1]}, \cite{[2]}  
based on backaction amplification \cite{[3]}. At the threshold of parametric instability, 
backaction-induced mechanical gain overcomes mechanical loss, resulting to an amplification 
process that leads to coherent phonon oscillations \cite{[1]}. Similar to stimulated emission of 
photon lasing in cavity with a gain medium, backaction amplification induces stimulated phonon 
lasing from the parametric instability's threshold \cite{[4]}. Both phonon lasing and amplification
are actively studied in optomechanics, and other fields, for technological purposes.

Single-phonon Fock state detection has been recently performed in \cite{[5]}, \cite{[6]}, and that paves 
a way to variety of quantum state engineering tasks including, quantum information processing 
\cite{[7]} and quantum entanglement of remote mechanical elements \cite{[8]}. The developpment of single 
phonon source reveals also a process towards technologies for precision sensing \cite{[9]}. Moreover, 
intensive researches are carried out in order to improve these achievements, for instance 
to lower the power need for phonon lasing. Such performance has been recently realized 
using $\mathcal{PT-}$symmetry optomechanics \cite{[10]}, and the polarization of light in 
coupled optomechanical devices \cite{[11]}. 

To use quantum signals, those involving only a few quanta, they need to be amplified and
must display some degree of purity to carry certain amount of information. This issue can be handled by 
characterizing the amplification process as the system gets closer to the phonon lasing threshold. These 
characteristics include phase-preserving amplification \cite{[12]}, \cite{[13]}, \cite{[14]}, the power gain and the number 
of added noise, and the gain-bandwidth product \cite{[15]}, \cite{[16]}, \cite{[17]}, \cite{[18]}. The 
purpose of a phase-preserving amplifier is to make a weak signal large, regardless of its phase, 
so that it can be perceived by devices unable to resolve the original signal, while sacrificing as little 
as possible in signal-to-noise ratio \cite{[14]}. Engineering amplifier having large 
power gain with quantum-limited added noise \cite{[17]}, \cite{[18]}, and without 
limitation on the gain-bandwidth product \cite{[17]} are useful for applications such as 
sensing \cite{[3]}, \cite{[9]} and quantum information processing \cite{[7]}. 
However, amplifying (or reaching phonon lasing threshold) with low-power consumption 
would provide practical advantage for amplifiers. Such low-power light amplifier has 
been recently realized, and has shown high gain at low-power operation and low quantum limit 
noise performance \cite{[19]}.

Recently, nonlinear position-modulated self-Kerr interaction has been engineered in COM, and it was 
found that it leads to an effective coupling that scales with the square  of the photon number \cite{[20]}. 
Such nonlinearity can be derived in the situation where Kerr nonlinear coefficient is modulated 
by the mechanical position. Such an interaction is possible in superconducting systems, cavity polaritonic systems, 
and atom-optical systems. It has been shown that this nonlinearity leads to a strong coupling, even for low driving strength. 
Moreover, these authors have demonstrated motional cooling, mode splitting and multistability for low-power red driving. 
Owing to these interesting nonlinear phenomena, our aim here is to perform low-power phonon laser at the blue sideband, 
resulting from low-power amplification induced by this nonlinearity. 
To understand this low-power operation, we have derived the cooperativity that highly
scales with the  photon number. This pushes the system fastly near to the
amplification's threshold. We have characterized the amplifier, that shows high gain and phase-preserving 
close to the phonon lasing threshold. This work opens up promising ways to develop low-power amplifiers based 
on the position-modulated self-Kerr interaction  in COMs and superconducting (electromechanical) microwave setups \cite{[21]}, 
\cite{[22]}. 

\section{Results}

\subsection{Hamiltonian and dynamical equations}
 
Position-modulated self-Kerr nonlinearity can be engineered in COMs \cite{[23]}, \cite{[24]}, \cite{[25]} or in the electromechanical microwave 
setups exhibiting a giant Kerr nonlinearity \cite{[21]}, \cite{[22]}. The idea of this engineering is based on the 
fact that, the Kerr nonlinear coefficient is modulated by the position of the mechanical resonator connected to the system.
Such nonlinear interaction, has been  recently investigated in \cite{[20]}. This has led to an effective coupling 
that scales with the square of the photon number. In the red sideband, 
this effective coupling leads to low powers required for motional cooling, the emergence of multistability, and 
other interesting nonlinear features. Owing to these exciting nonlinear effects, here we move to the blue sideband and 
investigate on phonon lasing and amplification phenomena. The Hamiltonian describing the generic system, with $\hbar=1$,
is given by,

\begin{equation}
H=-\Delta_{0}a^{\dag}a+\omega_{m}b^{\dag}b-g_{l}a^{\dag}a(b^{\dag
}+b)-g_{nl}a^{\dag}a^{\dag}aa(b^{\dag}+b)+E(a+a^{\dag}). \label{eq1}
\end{equation}

In this Hamiltonian, $a$ ($b$) is the annihilation bosonique operator for the intracavity field
(mechanical resonator),  $H_{l,int}=-g_{l}a^{\dag}a(b^{\dag}+b)$ and $H_{nl,int}=-g_{nl}a^{\dag}a^{\dag}aa(b^{\dag}+b)$ describe 
the linear and nonlinear interactions. The first two terms represent the cavity and mechanical 
free energy respectively, while the last term stand for the driving energy. The parameters $\omega_{m}$  and  
$\Delta_{0}=\omega_{p}-\omega_{cav}$ are the mechanical frequency of the resonator and  the detuning between the optical
drive ($\omega_{p}$) and the cavity eigenfrequency ($\omega_{cav}$).
The linear and nonlinear optomechanical couplings are denoted by $g_{l}$ and $g_{nl}$, respectively. 
The mechanical displacement is defined as $x=x_{ZPF}(b + b^{\dag})$, where 
$x_{ZPF}=\sqrt{\tfrac{\hbar}{2m\omega_m}}$ is the zero-point  fluctuation amplitude 
of the mechanical resonator, with $m$ its effective mass.
In the driving rotating frame, the Nonlinear Langevin Equations (NLEs), including cavity 
($\kappa$) and mechanical ($\gamma_m$) dissipations, can be derived as follows,
\begin{equation}
\left\{
\begin{array}{c}
\dot{a}=\left[i\left(\Delta_{0}+g_{l}(b^{\dag}
+b)+2g_{nl}(b^{\dag }+b)a^{\dag}a\right)-\frac{\kappa}
{2}\right]a-iE,\\
\dot{b}=-\left(i\omega_{m}+\frac{\gamma_{m}}{2}\right)b+ig_{l}a^{\dag
}a+ig_{nl}a^{\dag}a^{\dag}aa.
\end{array}
\right.  \label{eq2}
\end{equation}
Throughout the work, we assume the hierarchy of parameters $\gamma_{m}, g_{l}\ll \kappa \ll \omega_{m}$, 
similar to the experiments carried out in the resolved sideband regime \cite{[5]},\cite{[6]}. Our numerical 
and analytical investigations will be done at the sideband $\Delta_{0} = \omega_{m}$.
To get insight of the phonon lasing phenomenon, we linearize Eq. (\ref{eq2}) in the 
limit of large driving field. In this case, both intracavity field ($a$) and mechanical degrees of freedom ($b$) 
can be splitted into their average fields ($\alpha (t)$, $\beta (t)$) with some amount of 
fluctuations ($\delta\alpha (t)$, $\delta\beta (t)$) as follows,
\begin{equation}
\left\{
\begin{array}{c}
\delta\alpha(t)=a(t)-\alpha(t), \\
\delta\beta(t)=b(t)-\beta(t).
\end{array}
\right.  \label{eq3}
\end{equation}

Using Eq. (\ref{eq3}) in Eq. (\ref{eq2}), leads to the steady-state dynamics,

\begin{equation}
\left\{
\begin{array}{c}
\dot{\alpha}=\left(i\Delta-\frac{\kappa}{2}\right)\alpha+\sqrt{\kappa
}\alpha^{in},\\
\dot{\beta}=-\left(i\omega_{m}+\frac{\gamma_{m}}{2}\right)\beta
+ig_{l}\left\vert \alpha\right\vert ^{2}+ig_{nl}\left\vert \alpha\right\vert^{4},
\end{array}
\right.  \label{eq4}
\end{equation}
with the corresponding lowest order fluctuations dynamics, including noises,

\begin{equation}
\left\{
\begin{array}{c}
\delta\dot{\alpha}=\left(i\Delta-\frac{\kappa}{2}\right)  \delta\alpha
+i\chi\left(\delta\beta^{\ast}+\delta\beta\right)
+i\eta\left(\delta\alpha^{\ast}+\delta\alpha\right)
+\sqrt{\kappa}\delta\alpha^{in},\\
\delta\dot{\beta}=-\left(i\omega_{m}+\frac{\gamma_{m}}{2}\right)
\delta\beta+i\chi\left(\delta\alpha^{\ast}+\delta\alpha\right)
+\sqrt{\gamma_{m}}\delta\beta^{in}.
\end{array}
\right.  \label{eq5}
\end{equation}

In Eq. (\ref{eq4}) and Eq. (\ref{eq5}), we have set for convenience $-iE=\sqrt{\kappa}\alpha^{in}$ \cite{[18]} 
where the driving strength $\alpha^{in}$ is related to the 
input power $P_{in}$  as  $\alpha^{in}=\sqrt{\frac{P_{in}}{\hbar\omega_{p}}}$. We 
have defined the effective coupling as $\chi=\chi_{0}\alpha$ with $\chi_{0}=(g_{l}+2g_{nl}
|\alpha|^{2})$, $\eta=4g_{nl}|\alpha|^{2}\rm{Re}(\beta)$ and $\Delta=\Delta_{0}+2\chi
_{0}\rm{Re}(\beta)$. The noise operators are characterized 
by the \textit{noncommuting} properties, i.e., $\langle\delta\alpha^{in}(t)\rangle = 0$,
$\langle\delta\alpha^{in\dag}(t^{'})\delta\alpha^{in}(t)\rangle=n_{\alpha}\delta(t^{'}-t)$, and
$\langle\delta\alpha^{in}(t^{'})\delta\alpha^{in\dag}(t)\rangle=(n_{\alpha}+1)\delta(t^{'}-t)$ for the 
input field and $\langle\delta\beta^{in}(t)\rangle=0$, 
$\langle\delta\beta^{in\dag}(t^{'})\delta\beta^{in}(t)\rangle= n_{th}\delta(t^{'}-t)$, and 
$\langle\delta\beta^{in}(t^{'})\delta\beta^{in\dag}(t)\rangle =(n_{th}+1)\delta(t^{'}-t)$ for the 
thermal bath. The quantities $n_{\alpha}$ and $n_{th}$ are the equilibrium occupation numbers for the 
input field and mechanical oscillator, respectively. 

\subsection{Stability}

The steady-state solutions $\alpha_{s}$ and $\beta_{s}$  are derived from Eq. (\ref{eq4}), providing $\dot{\alpha}=0$  and $\dot{\beta}=0$.
These solutions are given by,
\begin{equation}
\left\{
\begin{array}{c}
\alpha_{s}=\frac{\sqrt{\kappa}\alpha^{in}}{\left(\frac{\kappa}{2}-i\left[\Delta_{0}+2\chi
_{0}\rm{Re}(\beta_{s})\right]\right)},\\
\beta_{s}= \left(g_{l}+g_{nl}\left\vert \alpha_{s}\right\vert ^{2}\right)\frac{\left\vert \alpha_{s}\right\vert ^{2}}{\omega_m-i\frac{\gamma_m}{2}}.
\end{array}
\right.  \label{eq4s}
\end{equation}
By setting the intracavity intensity as $I=|\alpha_{s}|^2$, one shows from Eq. (\ref{eq4s}) that it 
is solution of the following seventh-order polynomial equation,

\begin{equation}
I^{7}+a_{6}I^{6}+a_{5}I^{5}+a_{4}I^{4}+a_{3}I^{3}+a_{2}I^{2}+a_{1}I+a_{0}=0, \label{eq8}
\end{equation}
with $a_{6}=\frac{3g_{l}}{g_{nl}}$; $a_{5}=\frac{13g_{l}^{2}}{4g_{nl}^{2}}$; $a_{4}=\frac{3g_{l}^{3}+
\Delta\omega_{m}g_{nl}}{2g_{nl}^{3}}$; $a_{3}=\frac{g_{l}(g_{l}^{3}+3\Delta\omega_{m}g_{nl})}{4g_{nl}^{4}}$;
$a_{2}=\frac{\Delta\omega_{m}g_{l}^{2}}{4g_{nl}^{4}}$; $a_{1}=\frac{(
\frac{\kappa^{2}}{4}+\Delta^{2})\omega_{m}^{2}}{(  2g_{nl})  ^{4}}$; $a_{0}=-\frac{\kappa(\alpha^{in}\omega_{m})
^{2}}{(2g_{nl})^{4}}$. These steady state solutions are physically meaningless, unless they are stable. 
The stability can be given explicitly through Routh-Hurwitz criterion \cite{[25]}. However, we
analyze it here through linear stability theory, and confirm it with parametric instability threshold, since we are 
in the blue sideband regime. To this end, we start by writing Eq. (\ref{eq5}) in the follwing compact form,
\begin{equation}
\delta\dot{X}=M\delta X+\epsilon, \label{eq6}
\end{equation}
with $\delta X=(\delta\beta, \delta\beta^{\ast}, \delta\alpha, 
\delta\alpha^{\ast})^{T}$ and $\epsilon=(\sqrt{\gamma_{m}}\delta\beta^{in}, 
\sqrt{\gamma_{m}}\delta\beta^{in\ast}, \sqrt{\kappa
}\delta\alpha^{in}, \sqrt{\kappa}\delta\alpha^{in\ast})^{T}$. The
matrix $M$ is given by,

\begin{equation}
M=
\begin{bmatrix}
-(i\omega_{m}+\frac{\gamma_{m}}{2})  & 0 & i\chi & i\chi\\
0 & (i\omega_{m}-\frac{\gamma_{m}}{2})  & -i\chi & -i\chi\\
i\chi & i\chi & (i\tilde{\Delta}-\frac{\kappa}{2})  & i\eta\\
-i\chi & -i\chi & -i\eta & -(i\tilde{\Delta}+\frac{\kappa}{2})
\end{bmatrix},
\label{eq7}
\end{equation}
with $\tilde{\Delta}=\Delta+\eta$, and where $\alpha_{s}$ is assumed to be real ($\alpha_{s} \in \mathbb{R}$).
 
\begin{figure*}[tbh]
\centering
\par
\begin{center}
\resizebox{0.49\textwidth}{!}{
\includegraphics{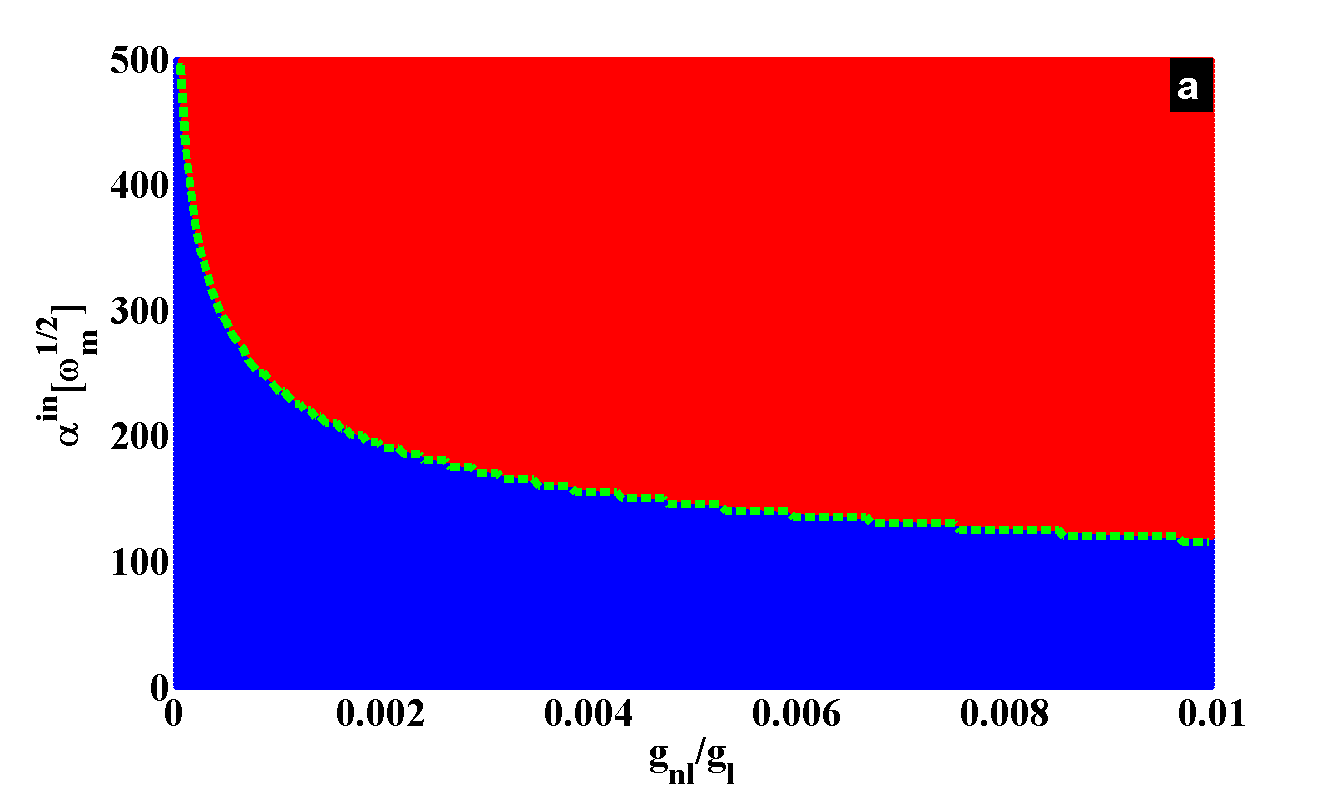}}
\resizebox{0.49\textwidth}{!}{
\includegraphics{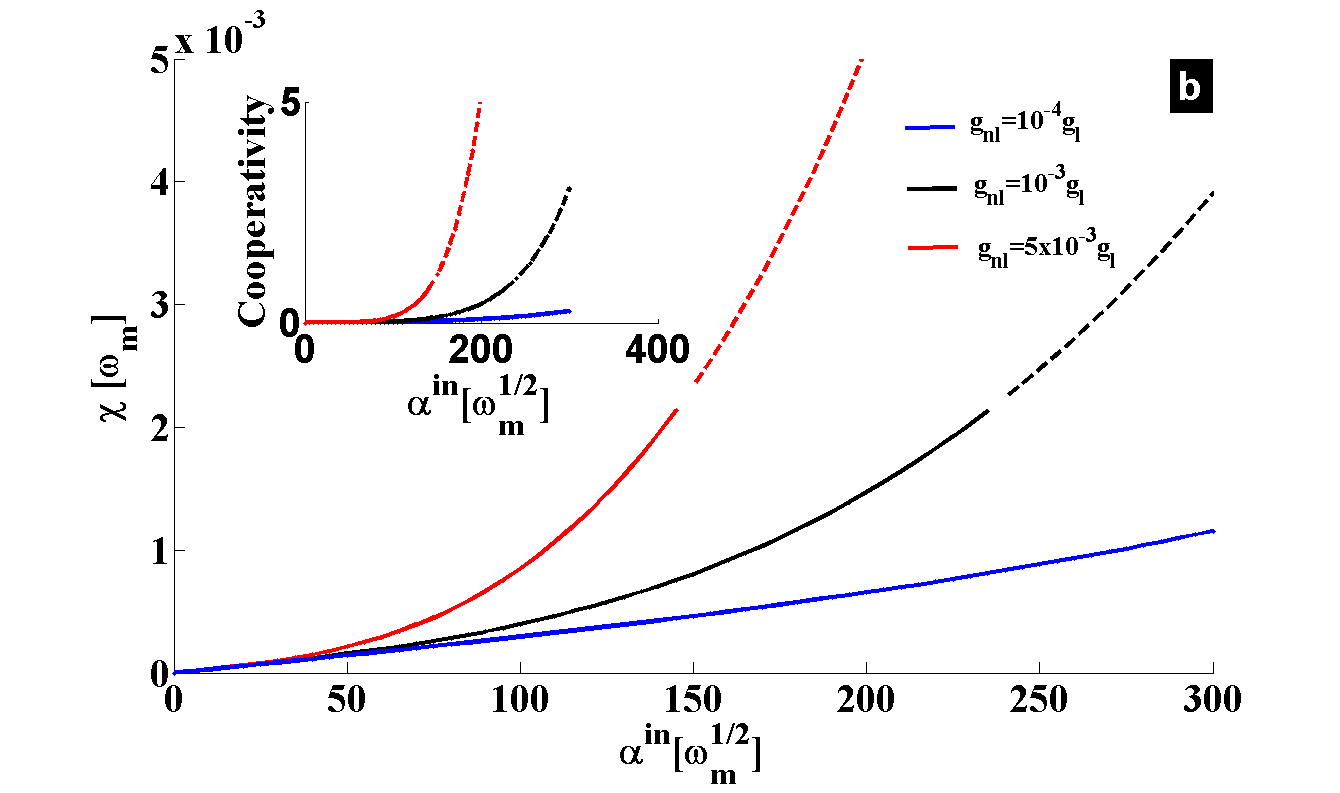}}
\resizebox{0.49\textwidth}{!}{
\includegraphics{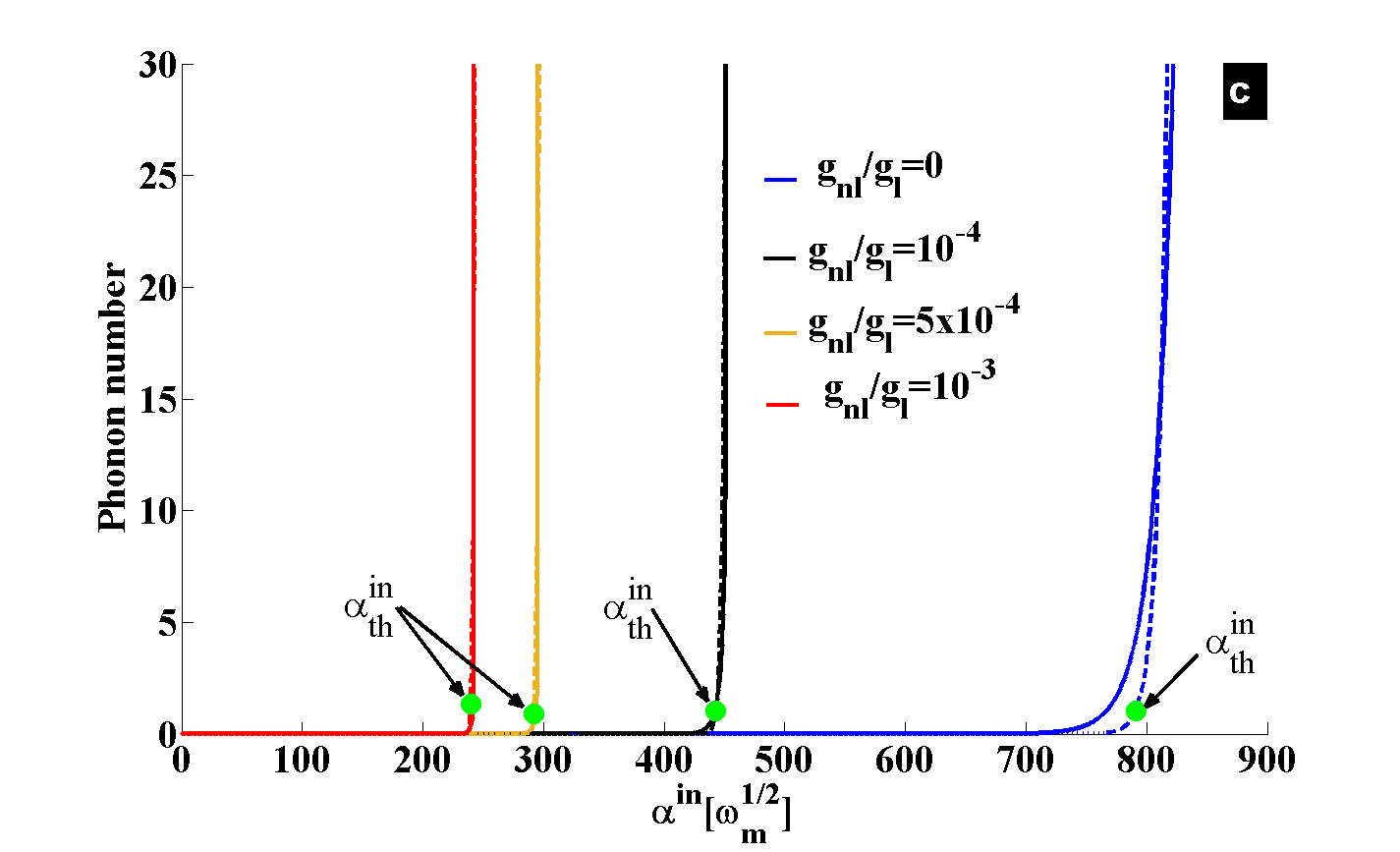}}
\resizebox{0.49\textwidth}{!}{
\includegraphics{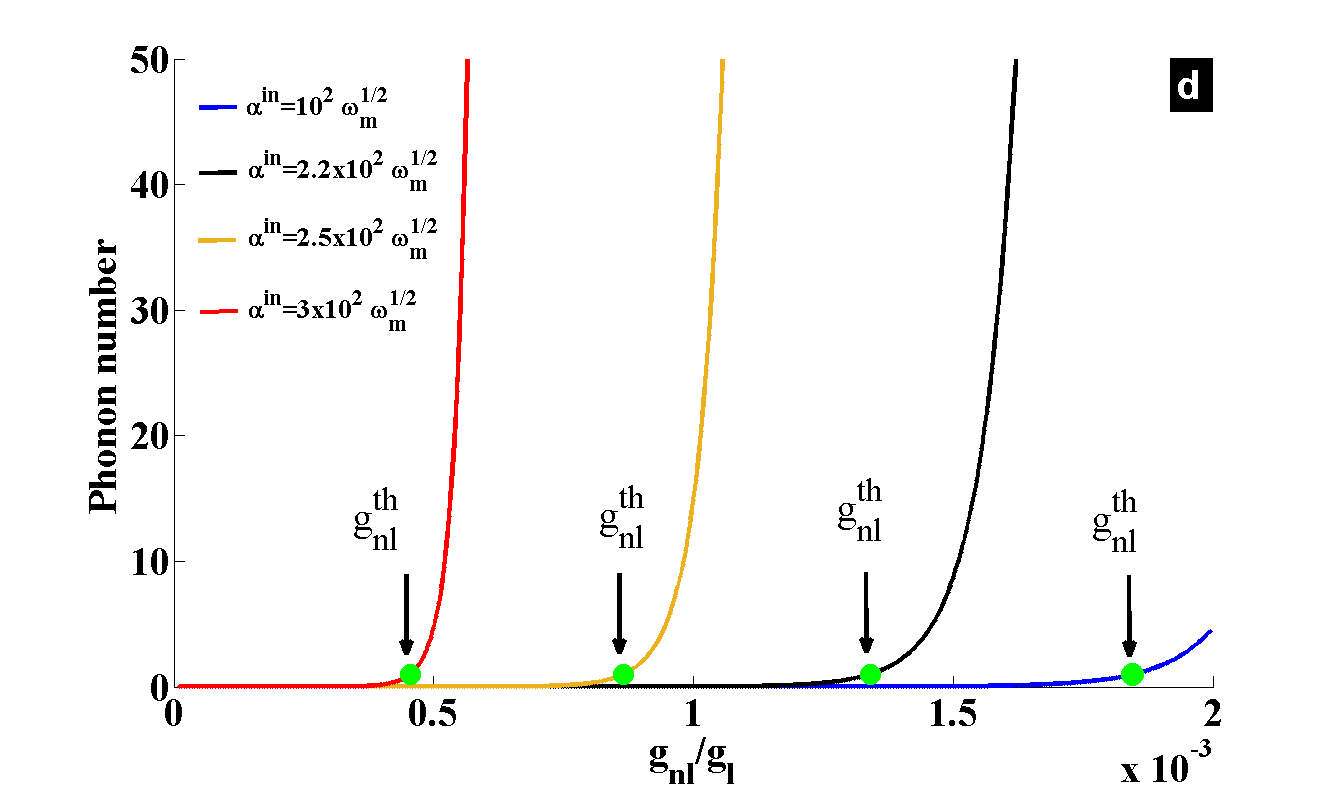}}
\end{center}
\caption{(a) Stability diagram. Blue (red) color is stable (unstable). The green dashed
curve shows the border between stable and unstable regions, and corresponds to the 
lasing threshold $\mathcal{C}=1$, with the cooperativity $\mathcal{C}=4\chi^{2}/(\gamma_m\kappa)$.
(b) Effective coupling $\chi$ versus $\alpha^{in}$ for different values of $g_{nl}$. Full (dashed) 
curves are stable (unstable). The inset shows the cooperativity $\mathcal{C}$ versus $\alpha^{in}$, 
for the corresponding values of $g_{nl}$. (c)  Phonon number versus $\alpha^{in}$. Full curves 
are from numerical simulation of classical version of Eq. (\ref{eq2}), while the dashed ones are the analytical approximation from 
$N\sim \rm{exp(-\gamma_{eff}t)}$. (d) Analytical (approximated) phonon
number versus $g_{nl}$. In (c) and (d), the different lasing thresholds are indicated by the green dots.  The used parameters are \cite{[20]},\cite{[22]}, $\gamma_m=10^{-3}\omega_m$, 
$\kappa=2\times10^{-2}\omega_m$, $g_{l}=2\times10^{-5}\omega_m$, and
$\Delta_{0}=\omega_m$.}
\label{fig:Fig1}
\end{figure*}

The system is stable if all the real part of eigenvalues 
($\lambda_{i=1..4}$) of the matrix $M$ are negative ($\rm{Re}(\lambda_{i=1..4})  <0$). This stability depends 
on steady-state solutions $\alpha_{s}$ and $\beta_{s}$,  and is shown in Fig. \ref{fig:Fig1}a.
The blue color in Fig. \ref{fig:Fig1}a depicts stable parameters space, while the red area shows the 
unstable zone. As the self-Kerr term ($g_{nl}$) is increasing, the system becomes unstable and the 
stability is limited for relatively weak driving strength $\alpha^{in}$. The  green dashed line in Fig. \ref{fig:Fig1}a holds for 
the condition $\mathcal{C}=1$, where $\mathcal{C}=4\chi^{2}/(\gamma_m\kappa)=\gamma_{opt}/\gamma_m$ is the cooperativity. 
This cooperativity, with $\gamma_{opt}$ being the optical damping, depicts the border between stable (linear)  and 
unstable (nonlinear) regimes. Moreover, $\mathcal{C}=1$ defines the threshold of the phonon lasing in the 
optomechanical blue sideband.
It results that the self-Kerr nonlinearity induces low power phonon lasing action, that can be understood by the enhancement of the 
effective coupling ($\chi$) shown in Fig. \ref{fig:Fig1}b. This coupling enhancement is a direct consequence 
of the fact that $\chi$ highly scales with the photon number in the presence of $g_{nl}$, resulting in a large cooperativity
 (see inset of Fig. \ref{fig:Fig1}b).  Hence, as $g_{nl}$ increases, the lasing threshold is shifted 
 towards weak driving strength $\alpha^{in}$, revealing low-power phonon lasing in our proposal.

\subsection{Low power phonon lasing} 

To evaluate the stimulated emission phonon number, we have simulated the classical equivalent of the nonlinear equation
given in Eq. (\ref{eq2}). This is valid for large enough photon number in the system ($|\alpha|^2\gg1$), so that both 
fluctuations from intracavity field and mechanical resonator can be neglected.
By Fast Fourier Transforming (FFT) the results, and collecting the mechanical peak at the resonance, 
 we have obtained the phonon number versus the driving $\alpha^{in}$, depicted in Fig. \ref{fig:Fig1}c. Full and dashed lines 
 in Fig. \ref{fig:Fig1}c show the numerical and analytical results, respectively. This 
 yields good agreement between numerical calculation and  analytics, that is detailed in section \ref{Meth}.
Lasing threshold depicted by green dots are given by the condition 
$\gamma_m=\gamma_{opt}$ (or $\mathcal{C}=1$). As $g_{nl}$ is increasing, Fig. \ref{fig:Fig1}c reveals a low 
driving strength $\alpha^{in}$ required for phonon lasing threshold. In Fig. \ref{fig:Fig1}d, 
we have represented the phonon number versus $g_{nl}$ for different $\alpha^{in}$. It results 
that $g_{nl}$ enhances stimulated emission of phonons. Indeed, for $g_{nl}\sim0$, there is no 
lasing up to $\alpha^{in}=3\times10^{2}\sqrt{\omega_m}$ in Fig. \ref{fig:Fig1}d. Therefore, by adding a small amount 
of nonlinearity to the system, the lasing threshold rises up and is shifted towards small $g_{nl}$ 
as $\alpha^{in}$ increases. Briefly speaking, the higher is the driving strength, smaller is the amount 
of nonlinearity ($g_{nl}$) to reach the lasing threshold, and vice-versa. This can be 
understood from the dynamics of $b(t)$ in Eq. (\ref{eq2}), showing that $g_{nl}$ supplies 
energy ($\propto g_{nl}|\alpha|^4$) to drive the mechanical resonator. This reveals why 
self-Kerr nonlinearity studied here, is quite interesting and different from 
Kerr nonlinearity \cite{[23]}, \cite{[24]}, \cite{[25]}, \cite{[21]}, quadratic nonlinearity \cite{[26]} 
and Duffing nonlinearities \cite{[27]} also studied in optomechanics.

\subsection{Low power amplification} 
\begin{figure*}[tbh]
\centering
\par
\begin{center}
\resizebox{0.49\textwidth}{!}{
\includegraphics{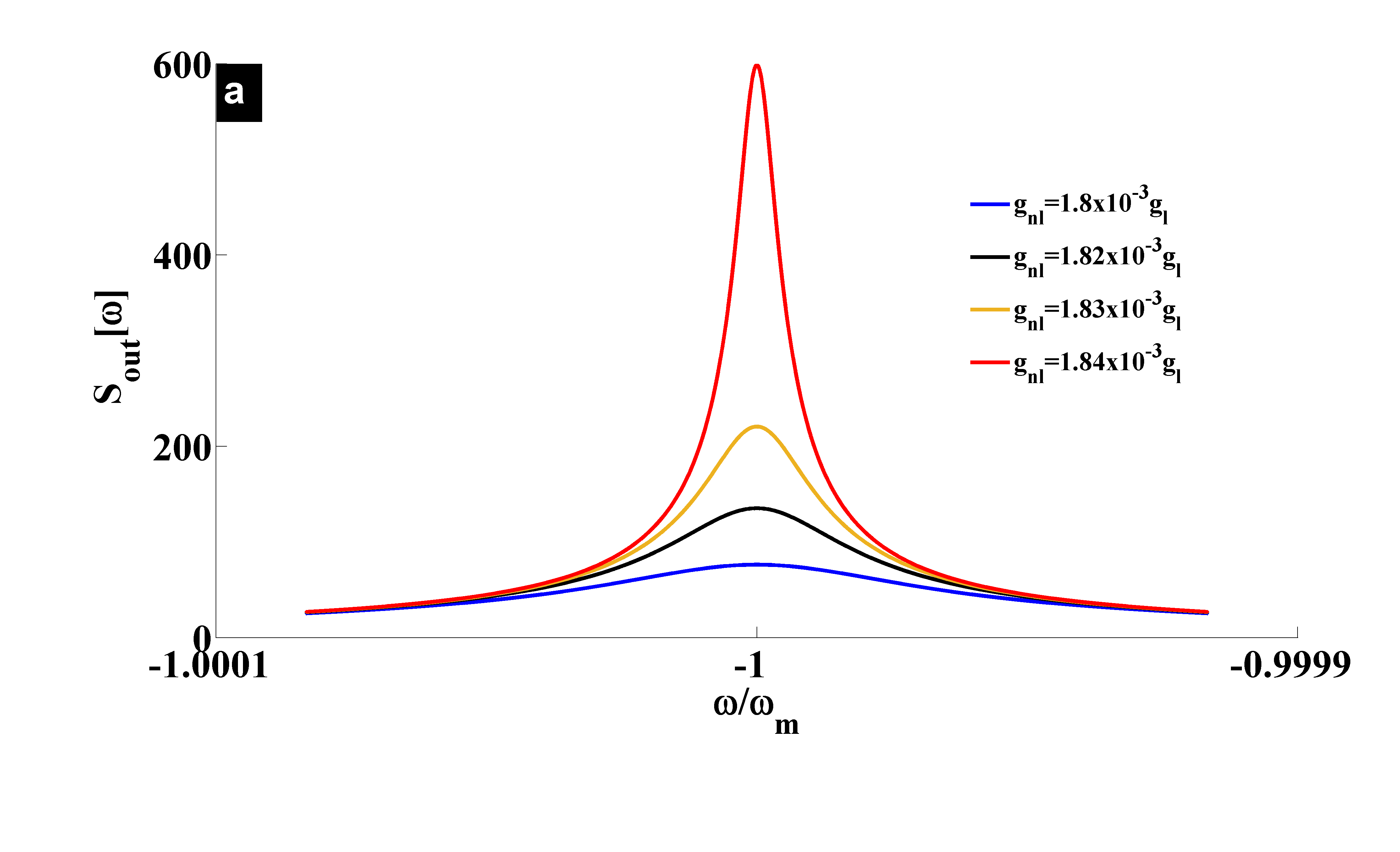}}
\resizebox{0.49\textwidth}{!}{
\includegraphics{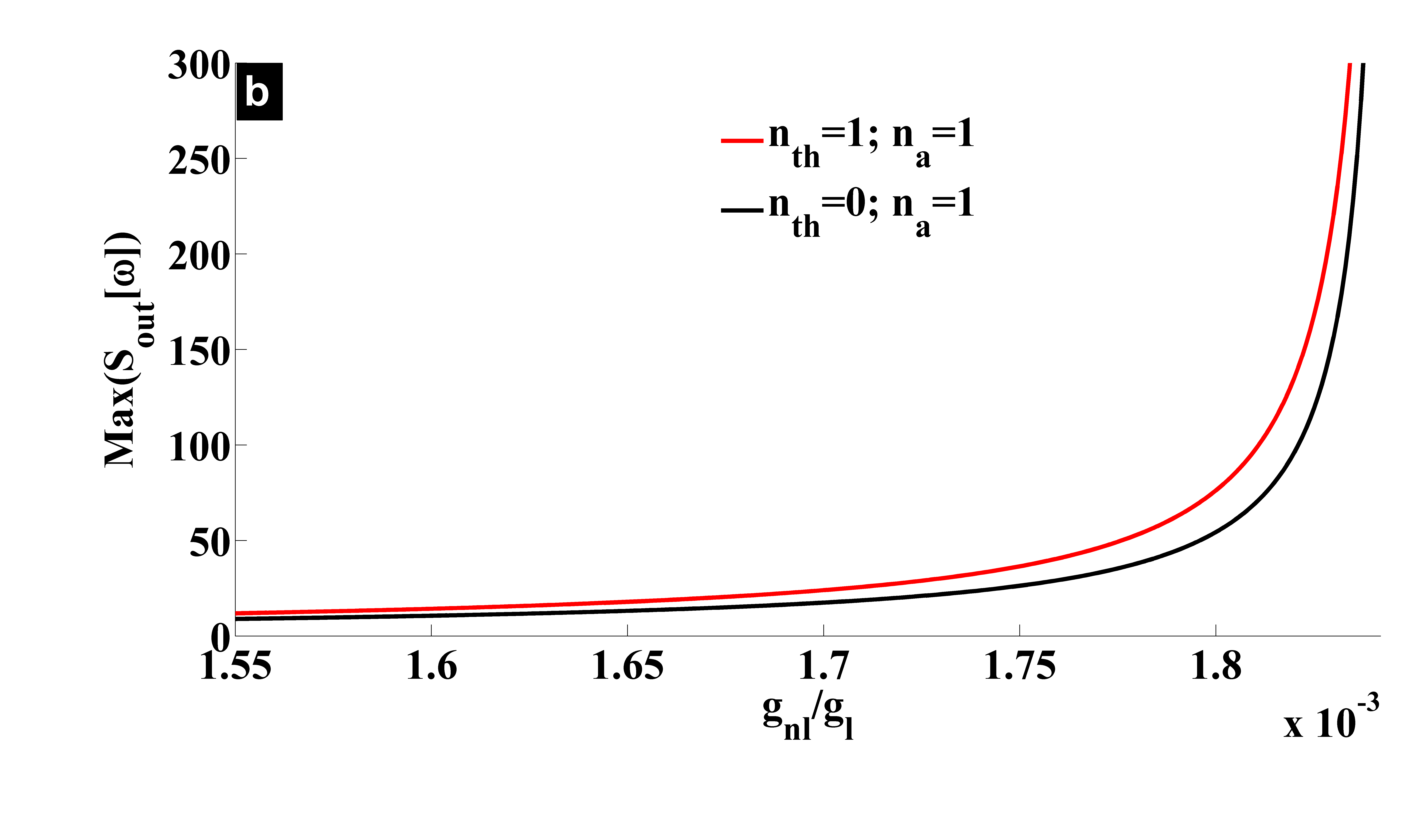}}
\resizebox{0.49\textwidth}{!}{
\includegraphics{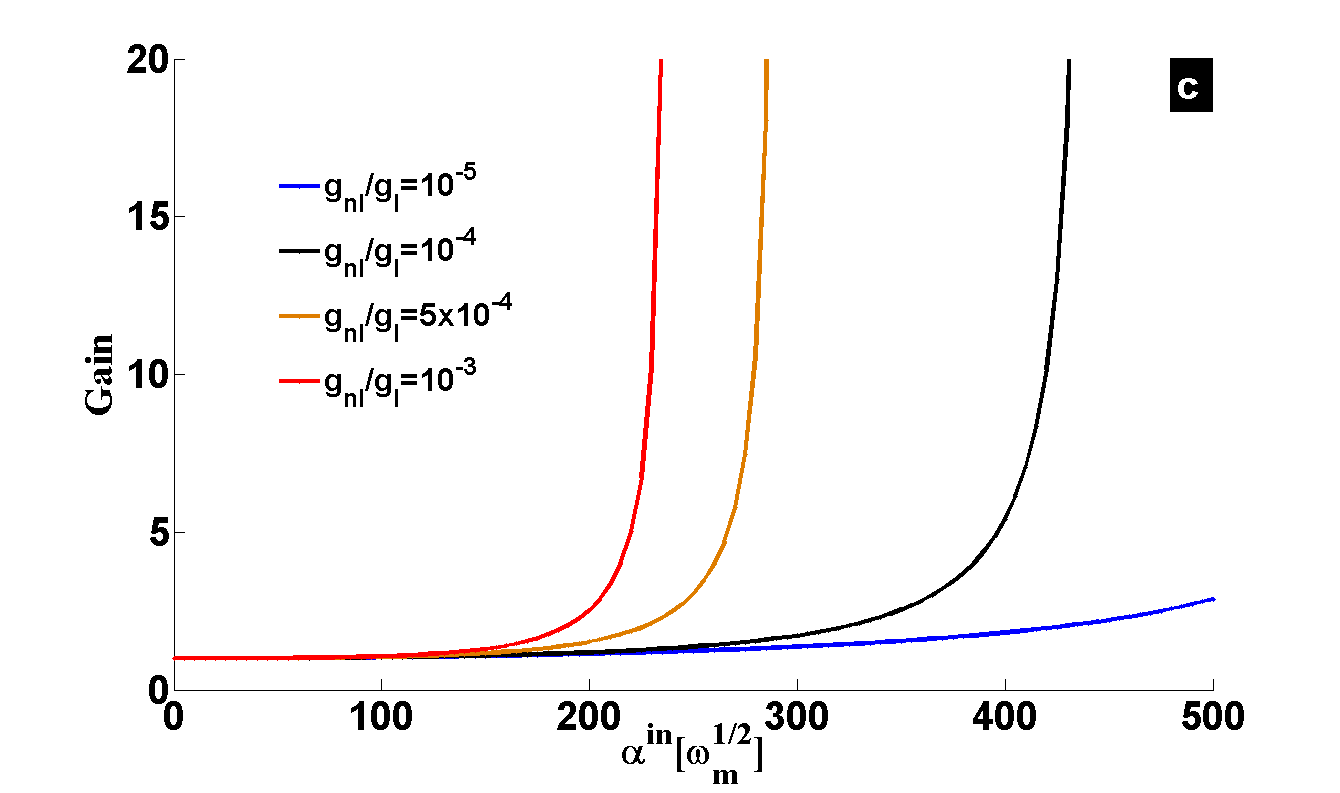}}
\resizebox{0.49\textwidth}{!}{
\includegraphics{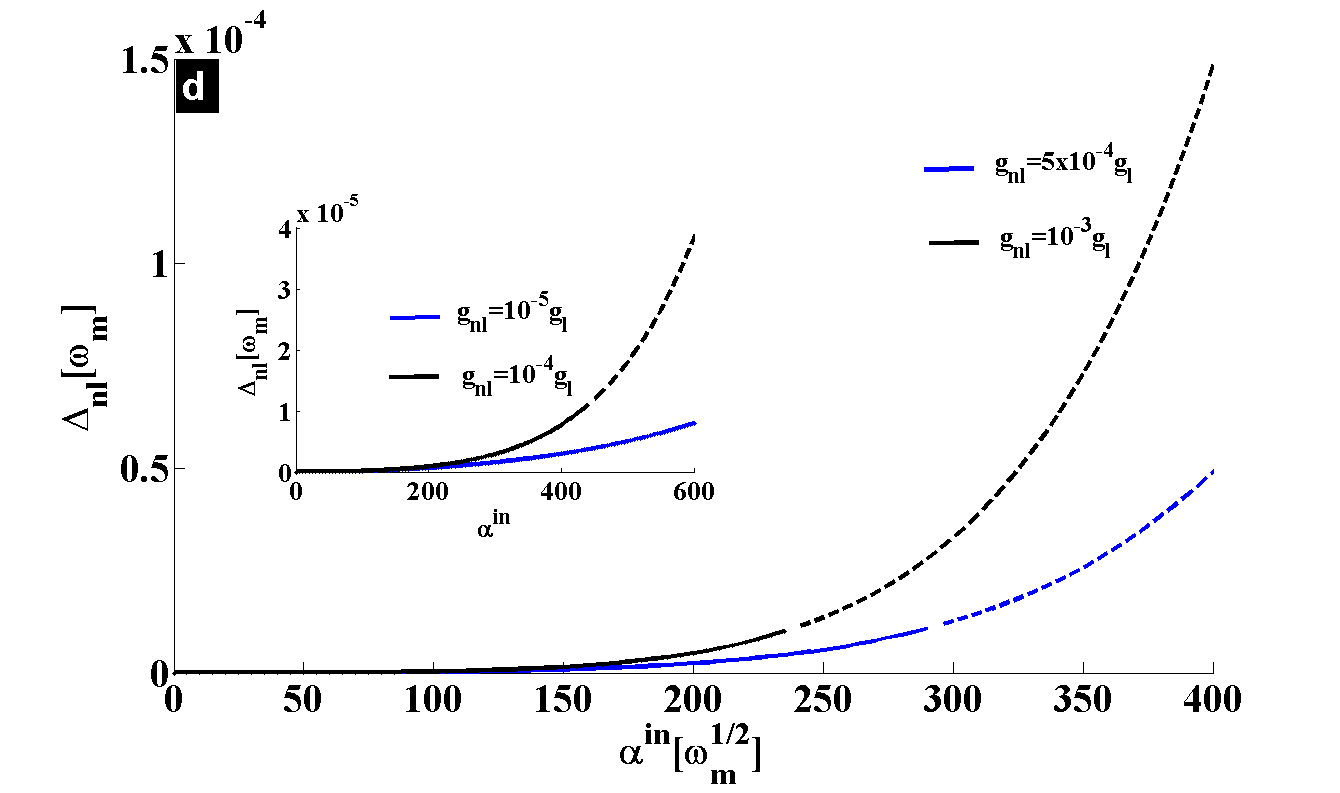}}
\end{center}
\caption{(a) Output noise power spectral density (PSD) for different values of $g_{nl}$. The magnitude of
$S_{out}=\sqrt{\langle\delta\alpha^{out\dag}\delta\alpha^{out}\rangle}$ reveals the 
amplitude of the mechanical resonator, detected at the output for a given frequency.  (b) Values of 
the output spectrum (in (a)) at the resonance ($\omega=-\Delta$). (c) Amplifier's gain 
(see Eq. (\ref{eq19})) versus $\alpha^{in}$, for different values of $g_{nl}$. (d) Cavity frequency shift 
$\Delta_{nl}=\tilde{\Delta}-\Delta_{0}$ versus $\alpha^{in}$. The driving strength in (a)-(b) is $\alpha^{in}=2\times10^{2}\sqrt{\omega_m}$, 
with $n_{th}=n_{\alpha}=1$ for (a). The other parameters are the same as in Fig. \ref{fig:Fig1}.}
\label{fig:Fig2}
\end{figure*}

As self-Kerr nonlinearity enhances low-power phonon lasing (Fig. \ref{fig:Fig1}c), this also 
reveals low-power amplification process in the system. The feature of the nondegenerate parametric amplifier here  
is to convert a pump mode photon into, one photon signal mode and  one idler phonon  mode. This leads
to the fact that weak incident signals are amplified, with a minimum possible added noise. 
Such amplification process can be seen from the linearization of the interaction Hamiltonian of our system.
Indeed, the interaction of our system is captured by the Hamiltonian,
\begin{equation}
H_{int}=H_{l,int}+H_{nl,int}=-g_{l}a^{\dag}a(b^{\dag}+b)-g_{nl}a^{\dag}a^{\dag}aa(b^{\dag}+b),
\end{equation}
which leads to its linearized form,
\begin{equation}
H_{int}^{lin}=-\chi(\delta\alpha^{\dag}\delta\beta^{\dag}+\delta\alpha\delta\beta)  
-\chi(\delta\alpha^{\dag}\delta\beta+\delta\alpha\delta\beta^{\dag}),  \label{eqb}
\end{equation}
after have used Eq. (\ref{eq3}). Furthermore, Eq. (\ref{eqb}) has been obtained by omitting static terms since they are 
taking into account in the frequency shift $\tilde{\Delta}$, and the higher order fluctuations terms 
have been neglected as being smaller than $\alpha_{s}$.
The second term on the right-hand side of Eq. (\ref{eqb}) stands for the counter rotating terms 
$H_{CR}=-\chi(\delta\alpha^{\dag}\delta\beta+\delta\alpha\delta\beta^{\dag})$ and can be 
neglected in the rotating wave approximation (RWA) \cite{[1]}. However, the first term 
($H_{R}=-\chi(\delta\alpha^{\dag}\delta\beta^{\dag}+\delta\alpha\delta\beta))$
describes a nondegenerate parametric amplifier, where a pump mode (photon) is converted 
into two quanta, one in the signal mode (photon), and the other in the idler (phonon). 
To characterize this amplifier, we neglect non resonant terms in Eq. (\ref{eq5}) and rewrite 
it in the RWA as, 
\begin{equation}
\left\{
\begin{array}{c}
\delta\dot{\alpha}=(i\Delta-\frac{\kappa}{2})
\delta\alpha+i\chi\delta\beta^{\dag}+\sqrt{\kappa}\delta\alpha^{in},\\
\dot{\delta\beta}^{\dag}=(i\omega_{m}-\frac{\gamma_{m}}{2})
\delta\beta^{\dag}-i\chi\delta\alpha+\sqrt{\gamma_{m}}\delta\beta^{in}.
\end{array}
\right.  \label{eqc}
\end{equation}
By solving Eq. (\ref{eqc}) in the Fourier space, together with the input-output relation, 
$\delta\alpha^{out}=\delta\alpha^{in}-\sqrt{\kappa}\delta\alpha$ \cite{[15]}, \cite{[18]}, we can evaluate the output 
field $\delta\alpha^{out}$. This output field is a key element to characterize the gain and added noise of the amplifier. 
More details on calculations leading to the gain and  added noise are reported in section \ref{Meth}, while 
specific results are shown in what follows. The input-output relation leads to the output field, 
\begin{equation}
 \delta\alpha^{out}=(1-\sqrt{\kappa}\chi_{eff}^{c})\delta\alpha^{in}-
 i\sqrt{\kappa}\eta_{c}\delta\beta^{in\dag}, \label{eq17}
\end{equation}
where 
\begin{equation}
\eta_{c}=\frac{\chi_{m}\chi_{c}\chi\sqrt{\gamma_{m}}}{1-\chi_{m}\chi_{c}\chi^{2}};  
\chi_{eff}^{c}=\frac{\chi_{c}\sqrt{\kappa}}{1-\chi_{m}\chi_{c}\chi^{2}},  \label{eq18}
\end{equation}
with the susceptibilities $\chi_{c}=\left[\frac{\kappa}{2}-i(\omega+\tilde{\Delta})\right]^{-1}$, and 
$\chi_{m}=\left[\frac{\gamma_{m}}{2}-i(\omega+\omega_{m})\right]^{-1}$. In Eq. (\ref{eq17}), the coefficient in front 
of the incident signal $\delta\alpha^{in}$ characterizes the amplification gain, while the one in front of the thermal 
noise informs on the added noise. These characteristics are deduced from the output noise power spectral density (PSD) 
defined as \cite{[17]}, \cite{[18]},
\begin{equation}
 S_{out}=\frac{1}{2}\left(\langle\delta\alpha^{out\dag}\delta\alpha^{out}+
\delta\alpha^{out}\delta\alpha^{out\dag}\rangle\right). \label{eql}
\end{equation}
The output PSD ($S_{out}$) is shown in Fig. \ref{fig:Fig2}a, and reveals an amplification process at the resonance 
$\omega=-\Delta$,  induced by the nonlinear term $g_{nl}$. For $\alpha^{in}=  2\times10^2\sqrt{\omega_m}$ 
and $g_{nl}\leq1\times10^{-3}$, there is no  amplification as shown in Fig. \ref{fig:Fig1}c. 
However, for $g_{nl}\approx1.85\times10^{-3}$, Fig. \ref{fig:Fig2}a clearly 
shows amplification, meaning  that $g_{nl}$ brings the system near the lasing threshold even for a weak driving strength. 
In order to appreciate this amplification, the output PSD has been plotted at 
the resonance $\omega=-\Delta$ (for $\alpha^{in}=  2\times10^2\sqrt{\omega_m}$) versus $g_{nl}$,  and shown in Fig. \ref{fig:Fig2}b. 
The red and black curves correspond to ($n_{\alpha}=1$, $n_{th}=1$) and ($n_{\alpha}=1$, $n_{th}=0$), 
respectively. It results that thermal noise has an impact on the amplification, 
revealing the effect of the added noise on the amplifier. This can be pointed out 
by evaluating both the power gain $\mathcal{G}(\omega)$ and the added noise $\mathcal{N}(\omega)$. 
The amplification of the signal  $\delta\alpha^{in}$ is measured through the gain 
$|1-\sqrt{\kappa}\chi_{eff}^{c}|^2$, which can be simplified (see section \ref{Meth}) at the resonance ($\omega=-\Delta$) as, 
\begin{equation}
\mathcal{G}=\left|\frac{\mathcal{C}+1}{1-\mathcal{C}}\right|^2.  \label{eq19}
\end{equation}
The noise performance of the amplifier is figured out through the input-referred added 
noise, defined as $\mathcal{N}(\omega)=\left(S_{out}-S_{in}\right)/\mathcal{G}^2$, 
where $S_{in}=\frac{1}{2}$ is the vacuum input noise driving the cavity. 
As shown in section \ref{Meth}, this expression reduces at the resonance to,
\begin{equation}
\mathcal{N}=\frac{4\mathcal{C}\left(n_{eff}+\frac{1}{2}\right)}{\left|\mathcal{C}+1\right|^2}, 
\label{eq20}
\end{equation}
where $n_{eff}=n_{th}+n_{\alpha}$ is the effective phonon number of the mechanical resonator.  Eq. (\ref{eq19}) shows 
that the gain exponentially grows as the system approaches the lasing threshold at a cooperativity 
$\mathcal{C}\rightarrow1$, revealing the amplification process. As $\mathcal{C}$ strongly 
scales with the self-Kerr term $g_{nl}$, it results an enhancement of amplification induced by 
$g_{nl}$ as shown in Fig. \ref{fig:Fig2}c. Furthermore, we have 
$\mathcal{N}\rightarrow \left(n_{eff}+\frac{1}{2}\right)$  for $\mathcal{C}\rightarrow1$. 
This reveals that the amplifier reaches the quantum limit for phase-preserving \cite{[13]}, \cite{[14]}
near the lasing threshold for $n_{eff}=0$, and if there are no additional loss channels. 
However, for non zero  effective (thermal) phonon number, $\mathcal{N}$ linearly increases with 
$n_{eff}$, degrading the signal amplification purity as depicted by the red curve in Fig. \ref{fig:Fig2}b. That  is not the case 
in the amplifier studied in \cite{[13]}, \cite{[17]} where the gain can take arbitrarily large values, and 
where any thermal noise contribution is suppressed for a large-gain limit \cite{[17]}. Despite of this, the 
practical advantage of our amplifier is its low power consumption due to the presence of $g_{nl}$.

\section{Methods} \label{Meth}
Fig. \ref{fig:Fig2}d shows that the cavity frequency shift ($\tilde{\Delta}-\Delta_{0}$) is weak in the linear regime that the 
intracavity field remains at the mechanical sideband $\tilde{\Delta} \sim \omega_m$. 
This allows us to get analytical expression of stimulated phonon number by introducing the slowly varying amplitude variables \cite{[28]}, 
$\delta\tilde{\alpha}=\delta\alpha\exp(\frac{\kappa}{2}-i\tilde{\Delta})t$ and $\delta\tilde{\beta}=\delta\beta\exp(i\omega_{m})t$. 
Using these variables in the cavity field equation in Eq. (\ref{eqc}) yields,
\begin{equation}
\delta\tilde{\alpha}=\int_{-\infty}^{t}i\chi\delta\tilde{\beta}^{\dag}e^{\frac{\kappa}{2}\tau}d\tau.  \label{eq9}
\end{equation}
As we are dealing with weak coupling regime in our analysis ($\kappa\gg\chi$, see Fig. \ref{fig:Fig1}b), 
we can adiabatically eliminate $\delta\alpha(t)$, then $\chi$  can be 
taken out of the integral Eq. (\ref{eq9}).  Moreover,  $\kappa\gg\gamma$ 
indicates that the evolution of $\delta\tilde{\beta(t)}$ is much slower that $\delta\alpha(t)$, 
meaning that $\delta\tilde{\beta(t)}$ can be considered as a constant term in Eq. (\ref{eq9}).
Under these conditions, the integration of Eq. (\ref{eq9}) yieds,
\begin{equation}
\delta\alpha=\frac{2i\chi}{\kappa}\delta\beta^{\dag}.  \label{eq10}
\end{equation}
By replacing Eq. (\ref{eq10}) in Eq. (\ref{eqc}), one gets
\begin{equation}
\delta\dot{\beta}^{\dag}=(i\omega_{m}-\frac{\gamma_{eff}}{2})\delta\beta^{\dag}, \label{eq11}
\end{equation}
where the effective damping is $\gamma_{eff}=\gamma_{m}-\gamma_{opt}$, with the optical damping 
$\gamma_{opt}=\frac{4\chi^{2}}{\kappa}$. The solution of $\mathcal{C}=1$ (or $\gamma_{eff}=0$),  with the 
cooperativity $\mathcal{C}=4\chi^{2}/(\gamma_m\kappa)$, gives the lasing threshold shown by the green dashed curve in 
Fig. \ref{fig:Fig1}a. The stimulated phonon number depicted in Fig. \ref{fig:Fig1} (c)-(d), is deduced from Eq. (\ref{eq11}), 
by evaluating $N=\langle\delta\beta^{\dag}\delta\beta\rangle=N_{0}e^{-\gamma_{eff}t}$, with $N_{0}=1$ 
the phonon number at $t=0$ \cite{[29]}.

The amplification shown in Fig. \ref{fig:Fig2} (a)-(b) are obtained by solving Eq. (\ref{eqc}) in the frequency domain, 
together with the input-output relation \cite{[15]}, \cite{[18]}. 
In the Fourier space, Eq. (\ref{eqc}) leads to,
\begin{equation}
\left\{
\begin{array}{c}
-i\omega\delta\alpha=(i\tilde{\Delta}-\frac{\kappa}{2})\delta\alpha+i\chi\delta\beta^{\dag}+\sqrt{\kappa}\delta\alpha^{in},\\
-i\omega\delta\beta^{\dag}=(i\omega_{m}-\frac{\gamma_{m}}{2})\delta\beta^{\dag}-i\chi\delta\alpha+\sqrt{\gamma_{m}}\delta\beta^{in\dag}.
\end{array}
\right.  \label{eq12}
\end{equation}

After some calculations, one obtains
\begin{equation}
\left\{
\begin{array}{c}
\delta\alpha(\omega)=\chi_{eff}^{c}\delta\alpha^{in}+i\eta_{c}\delta\beta^{in\dag},\\
\delta\beta^{\dag}(\omega)=\eta_{m}\delta\beta^{in\dag}-i\chi_{eff}^{m}\delta\alpha^{in},
\end{array}
\right.  \label{eq13}
\end{equation}
where
\begin{equation}
\left\{
\begin{array}{c}
\eta_{c}=\frac{\chi_{m}\chi_{c}\chi\sqrt{\gamma_{m}}}{1-\chi_{m}\chi_{c}\chi^{2}}; 
\chi_{eff}^{c}=\frac{\chi_{c}\sqrt{\kappa}}{1-\chi_{m}\chi_{c}\chi^{2}},\\
\eta_{m}=\frac{\chi_{m}\sqrt{\gamma_{m}}}{1-\chi_{m}\chi_{c}\chi^{2}}; 
\chi_{eff}^{m}=\frac{\chi_{m}\chi_{c}\chi\sqrt{\kappa}}{1-\chi_{m}\chi_{c}\chi^{2}},
\end{array}
\right.  \label{eq14}
\end{equation}
with the susceptibilities $\chi_{c}=\left[\frac{\kappa}{2}-i(\omega+\tilde{\Delta})\right]^{-1}$, and 
$\chi_{m}=\left[\frac{\gamma_{m}}{2}-i(\omega+\omega_{m})\right]^{-1}$.

Using the input-output relation, one gets the output field,
\begin{equation}
 \delta\alpha^{out}=(1-\sqrt{\kappa}\chi_{eff}^{c})\delta\alpha^{in}-
 i\sqrt{\kappa}\eta_{c}\delta\beta^{in\dag}, \label{eq15}
\end{equation}
which leads to the output PSD,
\begin{equation}
\begin{split}
S_{out} & =\frac{1}{2}\left(\langle\delta\alpha^{out\dag}\delta\alpha^{out}+
\delta\alpha^{out}\delta\alpha^{out\dag}\rangle\right) \\
     & =\kappa|\eta_{c}|^2(n_{th}+\frac{1}{2})+|1-\sqrt{\kappa}\chi_{eff}^{c}|^2 (n_{\alpha}+\frac{1}{2})\\
     &=\kappa|\eta_{c}|^2(n_{th}+\frac{1}{2})+\mathcal{G}(n_{\alpha}+\frac{1}{2}). \label{eq16}
\end{split}
\end{equation}

The amplifier is then characterized by the power gain, 
\begin{equation}
\mathcal{G}(\omega)=\left|1-\sqrt{\kappa}\chi_{eff}^{c}\right|^2=\left|\frac{1-\mathcal{C_{\omega}}
-2\left(1-\frac{2i}{\gamma_m}(\omega+\omega_m)\right)}{1-\mathcal{C_{\omega}}}\right|^2,
\end{equation}
with
$\mathcal{C_{\omega}}=\frac{4}{\gamma_{m}\kappa}\left(\chi^2+(\omega+\omega_m)(\omega+\tilde{\Delta})\right) 
-\frac{2i}{\gamma_{m}\kappa}\left(\gamma_m(\omega+\tilde{\Delta})+\kappa(\omega+\omega_m)\right)$.
The input-referred added noise quanta to the amplifier is defined as,
\begin{equation}
\begin{split}
\mathcal{N}(\omega) &=\left(S_{out}-S_{in}\right)/\mathcal{G}^2\\
 &=\frac{\kappa|\eta_{c}|^2}{\mathcal{G}}(n_{th}+\frac{1}{2})+n_{\alpha}\\
 &=\frac{4\mathcal{C_{\omega}}\left(n_{th}+\frac{1}{2}\right)}{\left|1-\mathcal{C_{\omega}}
-2\left(1-\frac{2i}{\gamma_m}(\omega+\omega_m)\right)\right|^2}+n_{\alpha},
\end{split}
\end{equation}
where $S_{out}$ is given in Eq. (\ref{eql}), and $S_{in}=\frac{1}{2}$ the vacuum input noise driving the cavity.
At the resonance ($\omega=-\Delta$), these amplifier's characteristic reduce to,
\begin{equation}
\mathcal{G}=\left|\frac{\mathcal{C}+1}{1-\mathcal{C}}\right|^2,
\end{equation}
and 
\begin{equation}
\mathcal{N}=\frac{4\mathcal{C}\left(n_{th}+\frac{1}{2}\right)}{\left|\mathcal{C}+1\right|^2}+n_{\alpha},
\end{equation}
which clearly show: (i) amplification near the lasing threshold 
($\mathcal{G}\rightarrow\infty$ for $\mathcal{C}\rightarrow 1$), and the fact that (ii) 
the amplifier reaches the quantum limit for a phase-preserving amplifier ($\mathcal{N}\rightarrow \frac{1}{2}$) for 
$\mathcal{C}\rightarrow 1$  and $n_{eff} \rightarrow0$,  
with $n_{eff}=n_{th}+n_{\alpha}$. 

\section{Discussion}
We have carried out investigation on position-modulated Kerr-type nonlinearity in blue 
sideband optomechanics. This is mainly focused on phonon lasing and amplification process. 
The amplification, which is based on the recorded output field, is shown to be as a manifestation of phonon lasing 
inside the cavity. Both phenomena are enhanced by this nonlinear term, through the condition $\mathcal{C}\rightarrow 1$. 
Indeed, a cooperativity equal to unity fulfills phonon lasing threshold requirement, large gain amplification, 
and condition of quantum limit for a phase-preserving amplifier. We have shown that these features happen for low-power strength, which 
is the figure of merit of the used nonlinear term, compared to those known so far \cite{[21]}, \cite{[23]}-\cite{[27]}.  
For accoustic excitations, using weak driving strength to reach phonon lasing, and large gain for 
the amplified phonon source having minimum added noise could be an interesting achievement. This can be 
useful for phonon information processing, and the position-modulated Kerr-type nonlinearity can be helpful, if $n_{eff}\rightarrow0$. 
For $n_{eff}\neq0$, one gets $\mathcal{N}\rightarrow n_{eff}+\frac{1}{2}$, 
showing how the effective (thermal) phonon number can impair the purity of the amplified signal. 
However, the practical benefit of low-power gain (phonon lasing) enhancement of such amplifier holds, 
as long as the self-Kerr nonlinearity $g_{nl}$ is involved.

\section*{Acknowledgments}

This work was supported by the European Commission FET OPEN H2020
project PHENOMEN-Grant Agreement No. 713450.

\end{document}